\documentclass[pre,aps,twocolumn,superscriptaddress,showpacs,floatfix]{revtex4-1}
\usepackage{amsmath}
\usepackage{amsbsy}
\usepackage{amssymb}
\usepackage[dvips]{graphicx}
\usepackage{color}

\begin{document}

\title{Approximated center-of-mass motion for systems of interacting particles with space and velocity dependent friction and anharmonic potential.}

\author{Alain Olivetti}
\email{alain.olivetti@unice.fr}
\affiliation{Cit\'e mixte du Parc Imp\'erial, 2 Avenue Paul Ar\`ene, 06050 Nice Cedex, France.}
\affiliation{Laboratoire J. A. Dieudonn\'e, UMR CNRS 6621,
Universit\'e de Nice-Sophia Antipolis, Parc Valrose, F-06108 Nice Cedex 02, France.}

\author{Guillaume Labeyrie}\author{Robin Kaiser}
\affiliation{Institut Non Lineaire de Nice, UMR 7335, 1361 route des Lucioles, F-06560 Valbonne, France}

\date{\today}

\begin{abstract}

  We study the center-of-mass motion in systems of trapped interacting particles with space and velocity dependent friction, and anharmonic traps. 
  Our approach, based on a dynamical ansatz assuming a fixed density profile, allows us to obtain information at once for a wide range of
  binary interactions and interaction strengths, at linear and non linear levels.
  Our findings are first tested on different simple models by comparison with direct numerical simulations.
  Then, we apply the method to characterize the motion of the center-of-mass of a magneto-optical trap, and its dependence on the number of 
  trapped atoms. Our predictions are compared with experiments performed on a large $Rb^{85}$ magneto-optical trap.

\end{abstract}

\pacs{05.10.-a, 45.50.-j, 05.45.-a, 05.20.-y}

\maketitle

\section{Introduction}

The study of low frequency modes often provides a convenient and non destructive tool to explore the properties of systems of trapped particles.
For instance, the measurement  of the breathing mode and the center-of-mass (c.o.m.) mode (also called sloshing mode or Kohn mode \cite{Kohn_1961}) frequencies allows to extract the Debye length \cite{Bonitz_2010}  and the particle charge \cite{Melzer_2001, Sheridan_2005}  in complex plasmas.
Similarly, in \cite{Moritz_2003} the authors characterize a trapped one-dimensional Bose gas by the frequency ratio of these two modes.

Numerous papers are devoted to the theoretical understanding of these low lying modes (see for instance \cite{Kohn_1961, Bonitz_2007, Henning_2008}). However, there is a need for theories covering more complex cases, such as particles experiencing a friction depending on space and/or velocity, or anharmonic traps, \ldots. For dusty plasma it has been shown that a negative friction may appear due to ion absorption by grains and create active particles \cite{Trigger_2003, Trigger_2003_2}. These features are also used to describe the so-called active Brownian particles in the context of plasma physics \cite{Dunkel_2004} or biological physics \cite{Mikhailov_1999, Erdmann_2005, Ebeling_2008}. In cold atoms experiments, a space dependent friction may have important dynamical consequences for the stability of magneto-optical traps \cite{Labeyrie_2006, Pohl_2006}. Finally, let us also mention that trapping anharmonicity may be essential to understand the dynamics of Bose-Einstein condensate \cite{Ott_2003}.
Our goal in this paper is to introduce a simple method able to deal qualitatively with the center-of-mass (c.o.m.) of trapped particles' systems, in situations where the friction may depend on space and/or velocity, and the trap may be anharmonic. This allows i) to avoid the simulations of the full system, which may be numerically costly ii) to emphasize the physical phenomena at play in the c.o.m motion in a simple qualitative analytical model.
We then compare our findings with experimental measurements on a Magneto-Optical Trap.
  
This paper is organized as follows. 
In section \ref{Sec: Scaling Ansatz} we introduce our model and the simplifying ansatz, which yields a prediction for the c.o.m. mode evolution. 
In section \ref{Sec: Test} we perform numerical tests, confronting the simplified theory with direct simulations of a one dimensional plasma system. 
Using different friction profiles and trapping potentials, we assess the validity of the method, and also emphasize its limitations.
Finally, in section \ref{Sec: MOT}, we study the c.o.m. mode for an atomic cloud in a Magneto-Optical Trap and predict that, for some parameters, the relaxation may change drastically as the number of atoms in the cloud is increased. This prediction is confirmed by experimental data obtained with a large magneto-optical trap of Rb$^{85}$.

\section{Equation of the Sloshing motion}\label{Sec: Scaling Ansatz}

Let us consider a system of $N$ particles confined by a trapping potential $\Phi(\mathbf{r})$, subject to binary interaction forces $\mathbf{F}_{bin}$.
To derive the motion of c.o.m. we consider the continuum limit of that system and write the first equation of the Bogolyubov-Born-Green-Kirkwood-Yvon (BBGKY) hierarchy plus a Fokker-Planck operator:

\begin{equation}\label{EQ: BBGKY + FP}
\frac{\partial f}{\partial t} + \boldsymbol\nabla_{\mathbf{r}}.\left(\mathbf{v}f\right) - \boldsymbol{\nabla}_{\mathbf{r}}\Phi.\boldsymbol\nabla_{\mathbf{v}}f+C[g]
=
\Delta_{\mathbf{v}}\left(Df\right)+\boldsymbol\nabla_{\mathbf{v}}.\left(\kappa\mathbf{v}f\right)
\end{equation}

with $f(\mathbf{r}, \mathbf{v}, t)$ the one-particle distribution, $D$ the diffusion coefficient and $\kappa(\mathbf{r}, \mathbf{v})$ the friction which may depend on the space and/or velocity coordinates. 
The interaction term is denoted $C[g]$ which is given by:

\begin{equation}
C[g](\mathbf{r}, \mathbf{v}, t) = \int \mathbf{F}_{bin}(\mathbf{r}, \mathbf{r}').\boldsymbol{\nabla}_{\mathbf{v}}g(\mathbf{r}, \mathbf{v}, \mathbf{r}', \mathbf{v}', t)d\mathbf{r}'d\mathbf{v}'
\end{equation}

and $g(\mathbf{r}, \mathbf{v}, \mathbf{r}', \mathbf{v}', t)$ the two-particles distribution.
Let us stress that Eq. \eqref{EQ: BBGKY + FP} is equivalent to Langevin equations for the particle dynamics because we have made no assumption about the unknown function $g$. However, solving the BBGKY hierarchy is as difficult as solving Langevin equations. 
In order to obtain our approximation of the c.o.m. mode we limit ourselves to the first moments of Eq. \eqref{EQ: BBGKY + FP}. 
Multiplying the equation \eqref{EQ: BBGKY + FP} by $r_j/N$ (resp. $v_j/N$), integrating over $d\mathbf{r}d\mathbf{v}$ and combining the results leads to:

\begin{equation}\label{EQ: Center-of-mass evolution}
\begin{array}{ll}
\displaystyle
\frac{\partial^2 \left\langle r_j\right\rangle_{f}}{\partial t^2} = & \displaystyle
-\left\langle \frac{\partial \Phi}{\partial r_j}(\mathbf{r})\right\rangle_{f} 
-\left\langle \kappa(\mathbf{r},\mathbf{v})v_j\right\rangle_{f}\vspace{2mm}\\
\displaystyle
&+\frac{1}{N}\int F_{bin}^j(\mathbf{r}, \mathbf{r}') g(\mathbf{r}, \mathbf{v}, \mathbf{r}', \mathbf{v}', t)d\mathbf{r}d\mathbf{v}d\mathbf{r}'d\mathbf{v}',
\end{array}
\end{equation}

where $j$ is a coordinate label, $F_{bin}^j$ the $j^{th}$ component of $\mathbf{F}_{bin}$ and we have set:

\begin{equation}
\left\langle \chi\right\rangle_{f} = \frac{1}{N}\int \chi(\mathbf{r},\mathbf{v}) f(\mathbf{r},\mathbf{v})d\mathbf{r}d\mathbf{v}.
\end{equation}

Thanks to the action reaction principle, the last term in \eqref{EQ: Center-of-mass evolution} vanishes, because the two-particles distribution is permutation invariant:

\begin{equation}
g(\mathbf{r}, \mathbf{v}, \mathbf{r}', \mathbf{v}', t) = g(\mathbf{r}', \mathbf{v}', \mathbf{r}, \mathbf{v}, t).
\end{equation}

We find here the classical result stating that the c.o.m. motion does not depend explicitly on the particles interaction. Note that this cancellation does not require any mean field hypothesis.
However, it is important to remark that for an anharmonic potential and/or non constant friction, the interaction appears implicitly in the distribution profile $f$, which is unknown.
Eq. \eqref{EQ: Center-of-mass evolution} is then not tractable.

To deal with the unknown distribution $f$, we drastically simplify the problem by considering a dynamics ansatz to only take into account the c.o.m. motion of the particles:
\begin{equation}\label{Eq: Ansatz}
f(\mathbf{r},\mathbf{v},t)= f_0(\varphi_t(\mathbf{r},\mathbf{v}))
\end{equation}
with
\begin{equation}
\varphi_t(\mathbf{r},\mathbf{v}) = \left( \mathbf{r}-\boldsymbol{\eta}(t), \mathbf{v}-\dot{\boldsymbol{\eta}}(t) \right)
\end{equation}
and $f_0$ is a stationary solution of Eq. \eqref{EQ: BBGKY + FP}. We also assume, without loss of generality, that $\langle \mathbf{r}\rangle_{f_0}=0$.
With this hypothesis all the time dependence in the dynamics is now included in the function $\boldsymbol{\eta}$ which is simply equal to $\langle \mathbf{r}\rangle_{f}$.
When the local mean velocity in the stationary state does not vanish, we expect that one should rather use:
\begin{equation}\label{Eq: Ansatz Part2}
\left\{
\begin{array}{ll}
\varphi_t(\mathbf{r},\mathbf{v}) = \left(\mathbf{r}-\boldsymbol{\eta}(t)\right., \left. \mathbf{u}_0\left(\mathbf{r}-\boldsymbol{\eta}(t)\right)+\mathbf{v}-\dot{\boldsymbol{\eta}}(t)\right)\\
\mathbf{u}_0(\mathbf{r}) = \int \mathbf{v}f_0(\mathbf{r},\mathbf{v})d\mathbf{v} / \int f_0(\mathbf{r},\mathbf{v})d\mathbf{v},
\end{array}
\right.
\end{equation}
In this article, we will stick to cases where $\mathbf{u}_0=0$.

Now, using the ansatz \eqref{Eq: Ansatz} and the Eq.\eqref{EQ: Center-of-mass evolution}, we easily obtain:
\begin{equation}\label{EQ: Sloshing motion}
\ddot{\eta}_j+
\langle \frac{\partial \Phi}{\partial r_j}(\mathbf{r}  + \boldsymbol{\eta})_{f_0} +
\left\langle \kappa\left( \mathbf{r} +\boldsymbol{\eta}, \mathbf{v}+\dot{\boldsymbol{\eta}}\right)
\left(v_j+\dot{\eta}_j\right)\right\rangle_{f_0}=0
\end{equation}
with $\eta_j$ the $j^{th}$ component of $\boldsymbol{\eta}$.
This result gives a generalization of the Kohn theorem \cite{Kohn_1961} where the whole system is spatially shifted.
Let us stress that in contrast with the constant friction case, even if it seems that the interactions do not appear, they are implicitly included in the shape of $f_0$, and thus they modify the evolution of $\mathbf{\eta}$.

\section{Numerical Test}\label{Sec: Test}

In this section we present numerical tests on the validity of Eq. \eqref{EQ: Sloshing motion} for different friction profiles and trapping potentials.

The continuous description used so far was convenient to develop the theory; on the other hand numerical simulations are easier going back to particles. Thus, we shall compare the theory~\eqref{EQ: Sloshing motion} to direct $N$-body simulations.
We integrate the $N$ Langevin equations using the Euler-Maruyama method. This method is adequate to compute average quantities when one is not interested in the exact trajectories of particles \cite{Talay_1990}.
Indeed, this scheme only uses one evaluation of the interaction forces at each time step and saves computation time in the most expensive part: the computation of binary forces, which is $O(N^2)$.
We first reach a stationary state, and then at time $t=0$ we spatially shift the whole system, and monitor the c.o.m. dynamics.

The benchmark system in this section is a trapped unidimensional plasma.
It consists of $N$ particles confined by an external trap.
We use a harmonic trap $\Phi(x)=(1/2)\omega^2x^2$, except in section \ref{Subsection: Anharmonic trap}.
The particles interact through a repulsive one dimensional Coulomb force.
In this simple case, the force depends only on the relative position of the particles: $\mathbf{F}_{bin}(x,x')=C\times\text{sgn}(x-x')$, and it allows to perform $N$-body numerical simulations with a high number of particles, without considering an approximate algorithm scheme, such as a tree-code \cite{Barnes_1986}, to compute binary forces.
Another advantage of our choice is that it allows to use an analytical approximation of the stationary space distribution profile $\rho_0$.
In the limit of strongly interacting/cold plasma with constant friction, the distribution profile is:
\begin{equation}\label{Eq: Plasma 1D stationnary profile}
\rho_0(x) = \left\{
\begin{array}{ll}
\frac{N}{2L_h}
,& \text{if } |x|\leq L_h\\
0,
&\text{elsewhere},
\end{array}
\right.
\end{equation}
as long as the typical size $L_h = NC/\omega^2$ is larger than $(D/\kappa_0\omega^2)^{1/2}$. Remark that the size of the system varies linearly with the number of particles and the interaction strength.\\
We will now compare the results of the analytical model (Eq.~\eqref{EQ: Sloshing motion}) as well as the analytical expression of the stationnary solution (Eq.~\eqref{Eq: Plasma 1D stationnary profile}) to  the numerical simulations. We will also see how a more precise knowledge of $f_0$ from simulations of the stationary state may increase the accuracy of
Eq.~\eqref{EQ: Sloshing motion}. Finally, we note that one expects a mean 
field description based on a Vlasov-Fokker-Planck equation to be accurate in 
the regimes explored numerically. Other tests would be needed to test the method in regimes where mean field descriptions break down.

\subsection{Space dependant friction}\label{Subsection: Space dependant friction}

We start by considering the following friction profile:
\begin{equation}\label{Eq: Space Friction profile}
\kappa(x,v) = \kappa(x) = \kappa_0\left(1 + \frac{|x|}{l}\right),
\end{equation}
with $l$ a typical size and $\kappa_0$ the friction value at $x=0$.

A priori, there is no reason that the stationary density $\rho_0$ keeps the shape given in Eq.~\eqref{Eq: Plasma 1D stationnary profile};
however our numerical simulations showed that Eq.~\eqref{Eq: Plasma 1D stationnary profile} remains a good approximation as long as we set parameters such that 

\begin{equation}
L_h\gg \frac{1}{\omega} \left(\frac{D}{\underset{|x|\leq L_h}{\min}\kappa(x)}\right)^{1/2}
\end{equation}

Using Eq.~\eqref{Eq: Plasma 1D stationnary profile} to compute the averages in Eq. \eqref{EQ: Sloshing motion}, we obtain a (non linear) 
equation for he c.o.m. motion:

\begin{equation}\label{Eq: Prediction Friction space variable}
\ddot{\eta} + \omega^2 \eta + \kappa_0\left(1+\frac{L_h}{2l}\right)\dot{\eta} + \frac{\kappa_0}{2lL_h}\dot{\eta}\eta^2 = 0. 
\end{equation}

\begin{figure}[!htpb]
\includegraphics[scale=0.35]{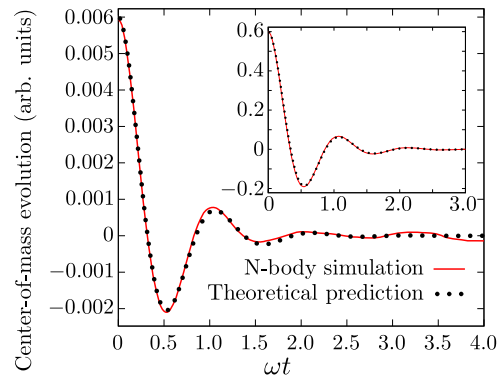}
\caption{(Color on line) Comparison between $N$-body simulations and theoretical predictions given by \eqref{Eq: Prediction Friction space variable} for small and large perturbations. 
Parameters are $\Delta t=0.001$, $N=10^5$, particles interaction $C=10^{-2}$, $l=1.0$, $\omega=17.8$, $\kappa_0=10.0$, $D=1.0$ and $\eta(0)\simeq 0.02\times L_h$.
We use the same parameters for the inset except $\eta(0)\simeq 2\times L_h$.\label{Fig: Friction space variable}}
\end{figure}

\begin{figure}[!ht]
\includegraphics[scale=0.35]{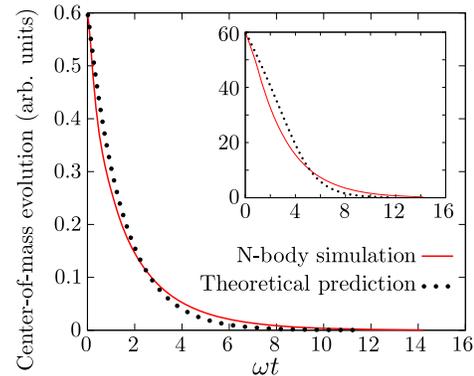}
\caption{(Color on line) Comparison between $N$-body simulations and theoretical predictions given by \eqref{Eq: Prediction Friction space variable} for small and large perturbations. 
Parameters are $\Delta t=0.001$, $N=10^5$, $C=1$, $l=1.0$, $\omega=17.8$, $\kappa_0=10.0$, $D=1.0$ and $\eta(0)\simeq 0.02\times L_h$.
We use the same parameters for the inset except $\eta(0)\simeq 2\times L_h$.\label{Fig: Friction space variable 2}}
\end{figure}

Figures \ref{Fig: Friction space variable} and \ref{Fig: Friction space variable 2} compare simulations and predictions with different perturbation amplitudes. A good agreement is obtained in strongly non-linear cases; it is even better when the local friction felt by the particles decreases. 
The agreement is less good however when the friction strongly varies (see figure \ref{Fig: Friction space variable 2} inset).
A similar behavior was observed for the breathing oscillation with space variable friction in~\cite{Olivetti_2011} and it
is closely related to the ansatz assumption that the profile does not change during the oscillation. To give a schematic view, let us consider two particles with the same velocity but with different positions. The first one being in a small friction area while the other one is highly damped.
Considering the same time step, the two particles do not cover the same distance.
Then the profile suffers some compression or dilatation, not included in the dynamical ansatz.
In summary, if the friction varies a lot and its values are not negligible with respect to the trapping constant, one may the assumptions behind \eqref{Eq: Ansatz} to be violated. 
The validity of our approach is thus related to the ratio friction/trapping, and we can summarize our findings as follows:
if $\max_{|x|<L_h}(\kappa(x))\ll \omega$, or the relative fluctuations of $\kappa(x)$ are small in the whole the system, then the dynamical ansatz assumptions are well satisfied. 


Beyond the comparison between predictions and simulations, figures \ref{Fig: Friction space variable} and \ref{Fig: Friction space variable 2} show an interesting phenomenon: when the size of the system increases, the c.o.m. motion changes drastically. 
Small systems undergo an underdamped relaxation, whereas large systems become overdamped.
Clearly, this is related to the ratio between the number of particles experiencing a large friction and those feeling a small friction. Eq.~\eqref{EQ: Sloshing motion} yields an approximate value for the threshold. 
Indeed by considering the linear expansion of Eq.~\eqref{EQ: Sloshing motion} we obtain the simple criterion:

\begin{equation}\label{Eq: Threshold prediction Over/Underdamped}
\left\langle \frac{\partial\kappa(\mathbf{r},\mathbf{v})v_j}{\partial v_j}\right\rangle_{f_0}^2
-
4\left\langle \frac{\partial^2 \Phi}{\partial r_j^2}(\mathbf{r}) \right\rangle_{f_0}^2
\left\{\begin{array}{ll}
< 0 & \Rightarrow \text{Underdamped}\\
> 0 &\Rightarrow \text{Overdamped}
\end{array}\right..
\end{equation}

It leads to the critical number of particles (or equivalently interaction strength in our specific example, see Eq.~\eqref{Eq: Space Friction profile}):

\begin{equation}
N_c = 2\frac{l\omega^2}{C}\left(2\frac{\omega}{\kappa}-1\right),
\end{equation}

and we conclude that for the same range of parameters, the behavior of c.o.m. of the plasma qualitatively changes when the number of particles increases.

\subsection{Velocity dependent friction}

We consider now another test case, where the friction varies linearly with the velocity:

\begin{equation}\label{Eq: Velocity Friction profile}
\kappa(x, v) = \kappa(v) = \kappa_0\frac{|v|}{v_0},
\end{equation}

with $v_0$ a typical velocity and $\kappa_0$ a typical friction.

We have no analytical expression for the stationary distribution $f_0$. However, we will see that a numerical estimation of $f_0$ is sufficient to obtain a quite good prediction of the c.o.m. motion.
In this case, we consider the discrete distribution of particles $f_0^N$ obtain from numerical simulation, which is more or less the one particle distribution:

\begin{equation}
f_0^N(x,v)=\sum_{i=1}^N\delta(x-x_i)\delta(v-v_i)\simeq f_0(x,v).
\end{equation}

Using $f_0^N$ to approximate $f_0$, Eq.~\eqref{EQ: Sloshing motion} leads to

\begin{equation}\label{Eq: Prediction Friction velocity variable}
\ddot{\eta} +\omega^2\eta +\frac{\kappa_0}{N}\sum_{i=1}^N  |v_i+\dot{\eta}|(v_i+\dot{\eta}) = 0.
\end{equation}

\begin{figure}[!htpb]
\includegraphics[scale=0.25]{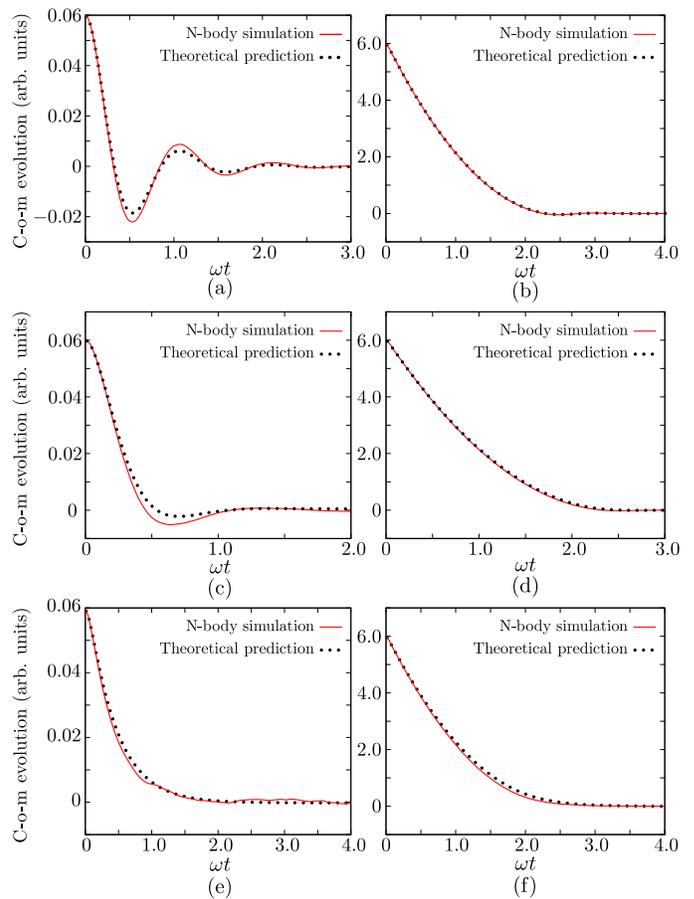}
\caption{(Color on line) Comparison between $N$-body simulations (straight line) and theoretical predictions (dotted line) given by \eqref{Eq: Prediction Friction velocity variable}. 
Figures (a), (c), (e) corresponds to small perturbation $\eta(0)\simeq 0.02\times L_h$. Figures (b), (d), (f) corresponds to large perturbation $\eta(0)\simeq 2\times L_h$.
The diffusion coefficient increases from top to bottom. (a)(b): $D=5$; (c)(d): $D=50$; (e)(f): $D=500$.
Other parameters are $\Delta t=0.001$, $N=10^5$, $C=10^{-2}$, $v_0=1.0$, $\omega=17.8$ and $\kappa_0=10.0$.
We continue to use parameter the $L_h$ because for these set of parameters the hypothesis of a strongly interacting or cold plasma is well satisfied.
\label{Fig: Friction velocity variable}}
\end{figure}

The evolution of Eq.~\eqref{Eq: Prediction Friction velocity variable} is computed using a fourth-order Runge-Kutta method and figure~\ref{Fig: Friction velocity variable} shows some simulations using \eqref{Eq: Velocity Friction profile} for different perturbation and diffusion coefficients.
Agreement is not perfect but very good results are obtained for large perturbations. This shows that the dynamical ansatz method is not limited to small perturbations.

In a similar manner as for the space-dependent friction case, we observe interesting features in figure~\ref{Fig: Friction velocity variable}. Varying the diffusion may change the dynamics of the c.o.m.
This phenomenon can be understood as follows: when $D$ increases, particles explore a larger region in phase space, including parts with larger velocity. Considering Eq.~\eqref{Eq: Velocity Friction profile}, the particles are then more damped and the global evolution changes from underdamped to overdamped.
Such a switching behavior between two qualitatively different evolutions has already been studied in different models with velocity dependent friction \cite{Mikhailov_1999, Erdmann_2005, Ebeling_2008}. 
In these papers the authors do not consider the c.o.m. motion, and in their cases the value of the diffusion coefficient implies a transition between a translation and a rotation mode. 
Nevertheless, we point out that the dynamical features are closely related to the nature of the friction and the shape of the one-particle distribution which depends on the diffusion coefficient $D$.

We now investigate a friction profile presenting large variations and a negative part:

\begin{equation}\label{Eq: Negative Friction}
\kappa(x,v)=\kappa(v)=-\kappa_0\left[1 - \left(\frac{v}{v_0}\right)^2\right],
\end{equation}
with $v_0$ a typical velocity and $\kappa_0$ the friction value at $v=0$. 
This friction profile is negative for $|v|<v_0$ and particles increase their energy in this region. With such a friction profile, the hypothesis underlying
\eqref{Eq: Ansatz} are expected to be completely violated.

\begin{figure}
\includegraphics[scale=0.25]{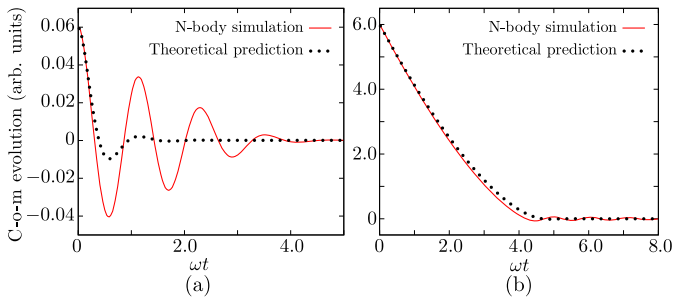}
\caption{(Color on line) 
Comparison between $N$-body simulations (straight line) and theoretical predictions (dotted line) given by the friction profile \eqref{Eq: Negative Friction}. 
(a): $\eta(0)\simeq 0.02\times L_h$; (b): $\eta(0)\simeq 2\times L_h$.
Other parameters are $\Delta t=0.001$, $N=10^5$, $C=10^{-2}$, $v_0=1.0$, $\omega=17.8$, $\kappa_0=10.0$ and $D=1.0$.
We continue to use parameter $L_h$ because for these set of parameters the hypothesis of strongly/cold interacting plasma is well satisfied.
\label{Fig: Negative Friction}}
\end{figure}

Figure~\ref{Fig: Negative Friction} shows the comparison between direct numerical simulations, and the prediction using 
\eqref{EQ: Sloshing motion} with the discrete density $f_0^N$.

In this case, as could be anticipated, the dynamical ansatz fails to predict the c.o.m. motion. 
Indeed, a negative friction induces some local dilation/compression which are not included in the dynamical ansatz.
For large perturbation we obtain a better result because the whole system starts in a positive friction region and the local dilation/compression effect become smaller. However, when the particles reach again negative friction a shift appears between prediction and simulation.

\subsection{Anharmonic trap}\label{Subsection: Anharmonic trap}

In this section we consider a one dimensional plasma with constant friction $\kappa$ in an  anharmonical trap. 
The trapping force used is:

\begin{equation}\label{Eq: Anharmonic Force}
\mathbf{F}_{trap}(x)=\frac{\omega^2}{1+4(\delta-\mu x)^2} - \frac{\omega^2}{1+4(\delta+\mu x)^2}
\end{equation}

with $\boldsymbol{\nabla}_{x}\Phi(x)=-\mathbf{F}_{trap}(x)$, $\delta<0$ and $\mu>0$. 

This kind of anharmonic trap appears for instance in some models of cold atoms in magneto-optical traps (see section \ref{Sec: MOT}).

\begin{figure}
\includegraphics[scale=0.25]{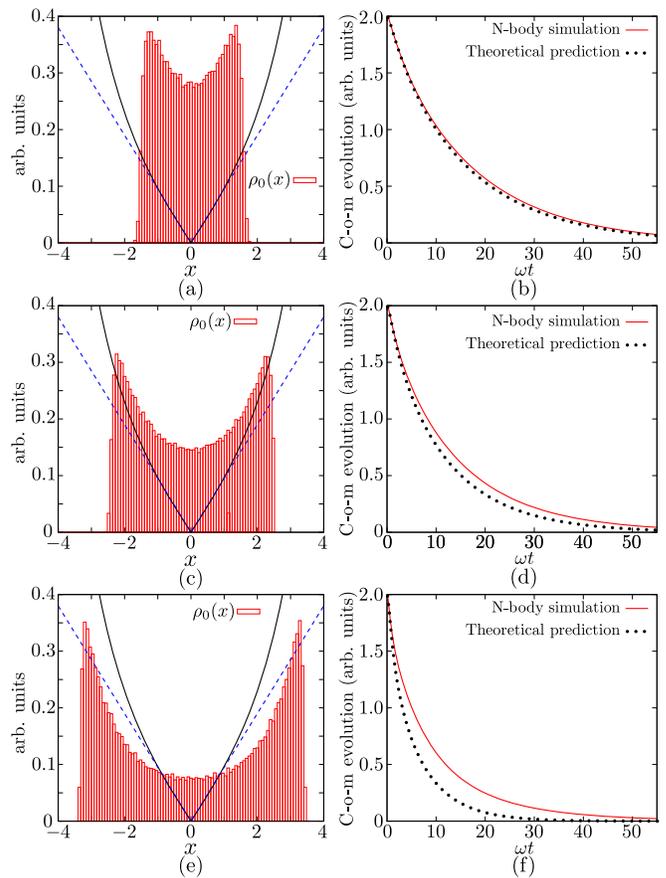}
\caption{(Color on line) 
Figures (a), (c), (e): Space stationary state from $N$-body simulations and absolute trapping force $|F_{trap}(x)|$ (straight line).
The absolute harmonic trapping force associated (dotted line) is included for reference purposes.
Figures (b), (d), (f): Comparison between $N$-body simulations (straight line) and theoretical prediction (dotted line) given by the trapping force~\eqref{Eq: Anharmonic Force}. 
The interaction strength increases from top to bottom: $C \in \{2.5, 5, 10\} (\times 10^{-5})$.
Others parameters are $\Delta t=0.001$, $N=10^5$, $\eta(0)=2.0$, $\delta=-6.0$, $\mu=1.0$ $\omega=17.8$, $\kappa=10.0$ and $D=0.1$.
\label{Fig: Anharmonic}}
\end{figure}

Figure~\ref{Fig: Anharmonic} shows a comparison between numerical simulation and prediction when the number of particles lying in the strongly anharmonic region becomes more and more important.
To highlight this aspect, the figure represents the absolute trapping force $|F_{trap}(x)|$ and the linear trapping force obtained by expanding $F_{trap}(x)$ around $x=0$.

Clearly the dynamical ansatz fails to describe the whole c.o.m. dynamics when to number of particles in the anharmonic region becomes too important.
We obtain the same results as for a negative velocity dependent friction: the dynamical ansatz hypothesis are not satisfied any more. 
As an example, \cite{Ott_2003} shows for a Bose-Einstein condensate that a non harmonic trap may lead to several new features. In their cases, they have identified several nonlinear effects due to anharmonicity, such as nonlinear mode mixing.

In conclusion, the dynamical ansatz method must be used with care when the number of particles in the anharmonic region is important in comparison with those in the harmonic one.
On the other hand, the dynamical ansatz yields satisfactory results for the c.o.m. motion in  a weakly anharmonic trap.

\section{Application: Large Magneto-optical trap}\label{Sec: MOT}

Recently it has been shown that increasing the number of atoms in a magneto-optical traps (MOT) may trigger dynamical instabilities, that are absent in smaller clouds \cite{Pohl_2006, Labeyrie_2006}. In this section, we apply the previous results to the case of the magneto-optical trap. We start by briefly introducing one of the simplest ways to model a MOT with Rb$^{85}$ atoms using the transition $F=3\rightarrow F'=4$ of the $D2$ line. We show that a the model naturally induces a space dependent friction, which leads to a qualitative change in the dynamics for the center-of-mass relaxation. Finally, we provide some experimental evidence confirming this prediction.

\subsection{Dynamics of the center-of-mass of a MOT}\label{Sec: MOTTh}

We use the so-called low-intensity Doppler model which is based on a velocity confinement due to the Doppler cooling and a spatial confinement due to the Zeeman effect \cite{Metcalf_1999}.
Let us stress that we neglect the sub-Doppler effect. We assume that the contribution of the sub-Doppler cooling does not change qualitatively the behavior of a large MOT with Rubidium atoms.
The average force $\mathbf{F}(\mathbf{r},\mathbf{v})$ acting on a single atom comes from the radiative pressure force:

\begin{equation}
\begin{array}{lcr}
F^i(r_i,v_i) &=&
\displaystyle 
\frac{\hbar k_L \Gamma}{2M}\frac{I_0}{I_{sat}}\frac{1}{1+\frac{4(\delta -\mu_ir_i-k_Lv_i)^2}{\Gamma^2}}\\
&-&
\displaystyle
\frac{\hbar k_L \Gamma}{2M}\frac{I_0}{I_{sat}} \frac{1}{1+\frac{4(\delta +\mu_ir_i + k_Lv_i)^2}{\Gamma^2}},
\end{array}
\end{equation}

with $F^i$ the $i^{th}$ component of $\textbf{F}$, 
$M$ the mass of Rb$^{85}$ atoms,
$I_0$ the laser intensity of lasers beams along the six directions, 
$I_{sat}$ the saturation intensity,
$\delta$ the detuning of the laser frequency $\omega_L$ with respect to the atomic resonance $\omega_A$ ($\delta=\omega_L-\omega_A<0$),
$\Gamma$ the natural linewidth of the transition used, 
$k_L$ the laser wave number and
$\mu_i r_i$ the Zeeman shift where $\mu_i$ depends on the applied magnetic field.

One usually considers the limit $k_Lv_i/\delta\ll 1$ and $\mu_i r_i/\delta\ll 1$, which allows to extract a friction and a trapping force. 
In the case of a large low temperature MOT, we assume that linearization in space is less reasonable than a linearization velocity.
The radiative pressure force becomes

\begin{equation}\label{Eq: Radiative Pressure Force Linearize}
\left\{\begin{array}{l}
F^i(r_i,v_i) = F_{trap}^i(r_i) - \kappa(r_i)v_i + \mathcal{O}\left(v_i\right)\\
F_{trap}^i(r_i)=\frac{\hbar k_L \Gamma}{2M}\frac{I_0}{I_{sat}}\left[
  \frac{\Gamma^2}{\Gamma^2+4(\delta -\mu_i r_i)^2}
- \frac{\Gamma^2}{\Gamma^2+4(\delta +\mu_i r_i)^2}
\right]\\
\kappa(r_i)=- \frac{\hbar k_L \Gamma}{2M}\frac{I_0}{I_{sat}}\left[
  \frac{8k_L\Gamma^2 (\delta -\mu_i r_i)}{\left(\Gamma^2+4(\delta -\mu_i r_i)^2\right)^2}
+ \frac{8k_L\Gamma^2 (\delta +\mu_i r_i)}{\left(\Gamma^2+4(\delta +\mu_i r_i)^2\right)^2}
\right].
\end{array}\right.
\end{equation}

In order to simplify our theoretical consideration and apply the dynamical ansatz method described in the section \ref{Sec: Scaling Ansatz}, 
we consider the symmetric approximation of the friction part of Eq.\eqref{Eq: Radiative Pressure Force Linearize} (in particular $\mu_i=\mu$):
\begin{equation}\label{Eq: Radiative Pressure Force Friction Approximate}
\begin{array}{ll}
\kappa(\mathbf{r})=&
\displaystyle
- \frac{\hbar k_L \Gamma}{2M}\frac{I_0}{I_{sat}} \times \\
&
\displaystyle
\left[
  \frac{8k_L\Gamma^2 (\delta -\mu|\mathbf{r}|)}{\left(\Gamma^2+4(\delta -\mu|\mathbf{r}|)^2\right)^2}
+ \frac{8k_L\Gamma^2 (\delta +\mu|\mathbf{r}|)}{\left(\Gamma^2+4(\delta +\mu|\mathbf{r}|)^2\right)^2}
\right],
\end{array}
\end{equation}

which coincides with expression \eqref{Eq: Radiative Pressure Force Linearize} along the axis.
This approximate friction \eqref{Eq: Radiative Pressure Force Friction Approximate} should preserve the important features of the system, albeit in a 
simplified way. 

It is commonly known in the cold-atoms community that this model is restricted to small MOT \cite{Sesko_1991}, \textit{i.e.} small number of particles. 
For large MOT we have to consider two other contributions: 
an attractive force $\mathbf{F}_A$ which comes from a screening effect, and a repulsive force which comes from multiple scattering.
In the small optical width region, these two forces satisfy \cite{Sesko_1991}:

\begin{equation}
\boldsymbol{\nabla}.\left[\mathbf{F}_A(\mathbf{r})\right] = -\frac{\sigma_L^2 I_0}{c} \rho(\mathbf{r})
\end{equation}

and 

\begin{equation}
\boldsymbol{\nabla}.\left[\mathbf{F}_R(\mathbf{r})\right] = \frac{\sigma_R\sigma_L I_0}{c} \rho(\mathbf{r}),
\end{equation}

where $\sigma_L$ (resp. $\sigma_R$) is the laser absorption (resp. atom scattering) cross section, $c$ the light velocity.
Assuming that $\mathbf{F}_A$ and $\mathbf{F}_R$ derive from a potential (which is not true for the former), we can use the Gauss theorem to consider these forces as a Coulombian binary interaction force:

\begin{equation}
\mathbf{F}_{bin}(\mathbf{r}, \mathbf{r}')=C\times\frac{\mathbf{r}-\mathbf{r}'}{|\mathbf{r}-\mathbf{r}'|^3}.
\end{equation}

The MOT is then described as a non-neutral plasma. 
When the temperature is low enough and the anharmonicity is weak, we obtain the same property as the toy-model used in our numerical test, \textit{i.e.} a stationary state:

\begin{equation}
\rho_0(\mathbf{r}) = \int f_0(\mathbf{r},\mathbf{v})d\mathbf{v} =
\left\{\begin{array}{cl}
\frac{3\omega^2}{C}& \text{, if } |\mathbf{r}|<L_h\\
0 & \text{, elsewhere}
\end{array},\right.
\end{equation}

with $L_h^3= NC/(4\pi \omega^2)$.

Exactly as in section \ref{Subsection: Space dependant friction}, we show that the friction profile given in Eq.~\eqref{Eq: Radiative Pressure Force Friction Approximate} leads to transitions between overdamped or underdamped relaxation when the number of particles change.
To predict the transitions between these two behaviors we linearize equation Eq.~\eqref{EQ: Sloshing motion}.
If we consider a perturbation along the $i^{th}$ direction, we obtain the condition:

\begin{equation}
\left\langle \kappa(\mathbf{r}) \right\rangle_{f_0}^2 - 4 \left\langle\frac{\partial^2 \Phi}{\partial r_i^2}(\mathbf{r}) \right\rangle_{f_0}  
\left\{
\begin{array}{ll}
<0 & \Rightarrow\text{Underdamped}\\
>0 & \Rightarrow\text{Overdamped}
\end{array}
\right.
\end{equation}

\begin{figure}
\includegraphics[scale = 0.40]{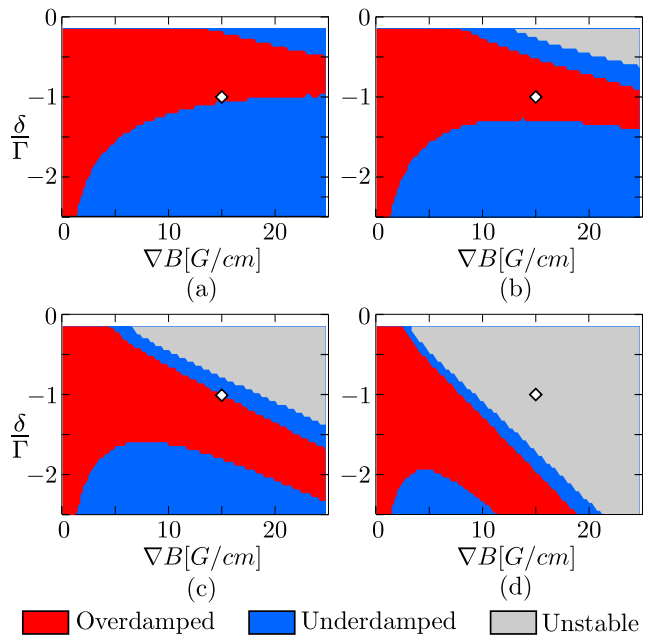}
\caption{(Color on line) Theoretical behavior of the c.o.m. mode
  relaxation in the parameters plane ($ \nabla B; \delta/\Gamma$) for
  different system sizes. The diamond shows the dynamical changes experienced by the c.o.m. motion as the MOT's size increases, at fixed parameter values ($15; -1$). (a):
  $L_h=0.5$~mm; (b): $L_h=1$~mm; (c): $L_h=2$~mm; (d): $L_h=4$~mm.  Other parameters are $k_L = 2\pi/\lambda$
  with $\lambda = 780\times10^{-9}$~m and $\mu = 2\pi\mu_0g\nabla B$
  with $\mu_0=2.1\times10^6$~G$^{-1}$ and $g=1.0$ an effective Land\'e
  g-factor of the transition used.
\label{Fig: Sloshing mode for MOT}}
\end{figure}

We represent on figure~\ref{Fig: Sloshing mode for MOT} the different behaviors, depending on the detuning $\delta$ and the magnetic field $\nabla B$ when the size of the system increases \textit{i.e.} when the number of particles increase.
It is possible to obtain three different behaviors depending on the parameter values: 
underdamped relaxation; overdamped relaxation; or the stationary state is unstable.
This latter possibility corresponds to $\langle \kappa(\mathbf{r})\rangle_{f_0}<0$. 
This implies that $\kappa(\mathbf{r})$ can be negative and that a lot of particles lies in those regions.
We will not discuss this case anymore since our numerical tests show that the method is unreliable when negative friction plays a role.
An important feature shown in figure~\ref{Fig: Sloshing mode for MOT} is that for the same value of detuning and magnetic field, 
it is possible to observe a modification of the relaxation dynamics by increasing the number of atoms. 

\begin{figure}
\includegraphics[scale = 0.35]{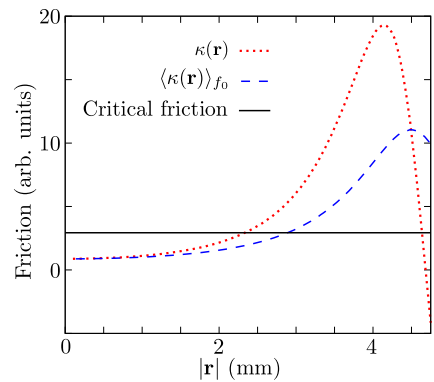}
\caption{(Color on line) Representation of the different friction depending on the size of the system for $\delta/\Gamma= -2.5$ and $\nabla B=15$~G/cm. 
Dotted line: friction profile $\kappa(\mathbf{r})$ given by \eqref{Eq: Radiative Pressure Force Friction Approximate};
dashed line: average friction $\langle \kappa(\mathbf{r})\rangle_{f_0}$;
straight line: critical friction considering the harmonic approximation of the trapping potential $\Phi$.
\label{Fig: Friction Profile MOT}}
\end{figure}

For example, increasing $L_h$ from $0.5$~mm to $4$~mm with $\delta/\Gamma=-2.5$ and $\nabla B = 15$~G/cm, the relaxation of the center-of-mass changes from underdamped to overdamped.
Considering the friction profile obtained for those parameters (see figure~\ref{Fig: Friction Profile MOT}) and using the same consideration as in section \ref{Sec: Test}, we are able to understand this behavior:
$\left\langle \kappa(\mathbf{r}) \right\rangle_{f_0}$ is just the average friction felt by the whole system; the threshold separating
 overdamped and underdamped is obtained by comparing this average friction with 
 $\left\langle \partial^2 \Phi /\partial r_i^2(\mathbf{r}) \right\rangle_{f_0}$.
Figure~\ref{Fig: Friction Profile MOT} represent these quantities for harmonic trap to simplify as much as possible the discussion.
When the size of the system is roughly between $2.5-3$~mm, the local friction at the edge of the system is higher than the critical friction, nevertheless the system is underdamped because the average friction stays below the threshold.
For $|\mathbf{r}|\geq 3$~mm the system becomes overdamped. For an even larger cloud, the friction decreases, the system may become underdamped again (not seen in the figure).

Let us stress that we used a simple model based on Doppler cooling to describe a MOT and many phenomena are not included in our study: sub-Doppler cooling is neglected (this can be important for the c.o.m. relaxation in small systems); the friction profile is assumed to be independent on the number of particles; the ``effective charge'' in the interaction force is taken as a constant.
We do not expect this model to predict exactly the transition between the different regimes; however, in some cases the Doppler model should be sufficient to describe qualitatively the system.
Note that \cite{Xu_2002} shows that atomic properties may dramatically changes the relaxation:
for alkaline-earth-metal atoms it is possible to observe underdamped oscillations \cite{Xu_2002}, while the same regime with alkali-metal atoms shows a strongly overdamped relaxation \cite{Steane_1991}.

\subsection{Experimental results}\label{Sec: MOTExp}

\begin{figure}
\includegraphics[scale = 0.30]{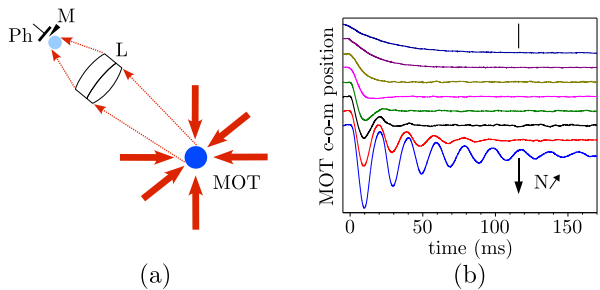}
\caption{(Color on line) Experimental observation of number-dependent MOT center-of-mass dynamics. (a) Detection setup. A fluorescence image of the MOT is formed on a photodiode (Ph) using lens (L). A mask (M) blocks half of the MOT image, rendering the photodiode signal sensitive to a lateral displacement of the MOT. (b) N-dependent dynamics. We record the sloshing motion of the cloud following a small displacement of the trap at $t = 0$. The different curves correspond to an increasing number of atoms from top to bottom.}
\label{Fig:damping}
\end{figure}

We present here some experimental evidence supporting the analysis developed in the previous section. Our experimental setup has been described elsewhere \cite{Labeyrie_2006}. We load a large magneto-optical trap containing up to 2$\times 10^{10}$ Rb$^{85}$ atoms from a dilute room-temperature vapor. The MOT employs 6 large (waist = 4 cm) independent laser beams, tuned slightly below (typically by $-3\Gamma$) the $F = 3 \rightarrow F' = 4$ transition of the $D2$ line. An additional repumping beam is also applied to the atoms, whose intensity is used to control the number of atoms in the $F = 3$ hyperfine level. To study the dynamics of the MOT's center-of-mass, we image the fluorescence light of the cloud on a photodiode, with a mask blocking half of the MOT's image (see figure~\ref{Fig:damping}(a)). This setup is thus sensitive to any displacement of the center-of-mass of the cloud in the plane orthogonal to the line of sight of the imaging optics.

Figure~\ref{Fig:damping}(b) shows the measured impact of the atom number on the dynamics of the MOT's center-of-mass. 
Using an offset magnetic field, we displace the MOT's center-of-mass. When this offset field is switched off at $t=0$, the MOT is free to evolve in the initial trapping conditions and the MOT then returns to its equilibrium position. As shown on figure~\ref{Fig:damping}(b), for low atom number the dynamics of the center of mass is overdamped, typical for standard MOTs. However we observe a clear transition from an overdamped behavior at low atom number, to an underdamped one when the atom number is large. This cross-over is heralding the instability regime observed in \cite{Labeyrie_2006}. This transition to an underdamped motion of the center of mass of the MOT before reaching the instability regime is similar to the narrow region indicated in blue in figure~\ref{Fig: Sloshing mode for MOT} separating the overdamped region from the instability region.

\section{Conclusion}

In the present work we have introduced a dynamical ansatz to obtain an approximate evolution of the center-of-mass mode considering a trapped system of interacting particles, assuming a global translation of the whole system.
This approach allows to describe numerous problems with arbitrary perturbation amplitudes of the center-of-mass mode.
This includes systems with space and/or velocity dependent friction as well as anharmonic trapping potential.
We have confronted the predictions of the simplified approach with direct $N$-body simulations of Langevin equations, considering a one dimensional plasma with different friction and trap profiles as test case.
The main conclusion is that the agreement is satisfactory as long as the hypotheses underlying the ansatz are well enough satisfied.
Using this approach on a model for a magneto-optical trap, we predict transitions between overdamped and underdamped motion for the center of mass of the cloud, as the number of trapped atoms increases. Finally, we provide some experimental evidence for such a transition, thus confirming some of the predictions made.

Note that we assume in this work a constant diffusion coefficient and an external force which derives from a potential.
However it is straightforward to extend the dynamical ansatz method. For instance in the case of a space and/or velocity dependent diffusion or with rotational forces.
Finally a combination of the dynamical ansatz introduced in this paper and the scaling ansatz method detailed in \cite{Olivetti_2009, Olivetti_2011} may be useful to describe simultaneously the center-of-mass motion and the breathing mode oscillation \cite{Guery_Odelin_2002}.

\acknowledgments 
This work is partially supported by the F\'ed\'eration W. D\"oblin (FR 2800). Alain Olivetti is grateful to J. Barr\'e, D. Broizat and C. Garc\'ia for fruitful discussions.

\end{document}